\documentclass[final,5p,times,twocolumn]{elsarticle}

\usepackage{amssymb}
\usepackage{here}

\biboptions{sort&compress}

\journal{Physica C}

\begin{document}

\begin{frontmatter}



\title{Unconventional superconductivity in electron-doped layered metal nitride halides $M$N$X$ ($M$ = Ti, Zr, Hf; $X$ = Cl, Br, I)}


\author[kas]{Yuichi Kasahara}
\ead{ykasahara@scphys.kyoto-u.ac.jp}
\author[kur]{Kazuhiko Kuroki}
\ead{kuroki@phys.sci.osaka-u.ac.jp}
\author[yam]{Shoji Yamanaka\corref{cor1}}
\ead{syamana@hiroshima-u.ac.jp}
\author[tag]{Yasujiro Taguchi\corref{cor2}}
\ead{y-taguchi@riken.jp}

\cortext[cor1]{Corresponding author}
\cortext[cor2]{Principal corresponding author}

\address[kas]{Department of Physics, Kyoto University, Kyoto 606-8502, Japan}
\address[kur]{Department of Physics, Osaka University, Toyonaka, Osaka 560-0043, Japan}
\address[yam]{Department of Applied Chemistry, Graduate School of Engineering, Hiroshima University, Higashi-Hiroshima, Hiroshima 739-8527, Japan}
\address[tag]{RIKEN Center for Emergent Matter Science (CEMS), Wako 351-0198, Japan}

\begin{abstract}
In this review, we present a comprehensive overview of superconductivity 
in electron-doped 
metal nitride halides $M$N$X$ ($M$ = Ti, Zr, Hf; $X$ = Cl, Br, I) with layered crystal structure and two-dimensional electronic states. 
The parent compounds 
are band insulators 
with no discernible long-range ordered state. 
Upon doping tiny amount of electrons, 
superconductivity emerges with 
several anomalous 
features 
beyond 
the conventional electron-phonon mechanism, 
which 
stimulate theoretical investigations. We will 
discuss 
experimental and theoretical 
results reported thus far 
and compare the electron-doped layered nitride superconductors with other superconductors.  
\end{abstract}


\end{frontmatter}


\section{Introduction}
Doping charge carriers into insulating materials is regarded as a key strategy for creating novel superconducting 
materials. Most fascinating systems that have been studied extensively thus far are high-$T_c$ cuprates \cite{Bednorz86} with layered crystal structure and strong electron correlation. The latter is believed to govern the physical properties in the normal and superconducting 
states, and indeed unconventional pairing mechanisms have been proposed. There is another class of superconductors induced by carrier doping, such as SrTiO$_3$ \cite{Koonce67}, diamond \cite{Ekimov04}, and silicon, which are semiconductors or band insulators. In most cases, such systems have three-dimensional (3D) crystal and electronic structure, and superconductivity can be understood in the framework of the conventional electron-phonon 
mechanism based on the 
Bardeen-Cooper-Schrieffer (BCS) 
theory. Here we discuss superconductivity in electron-doped layered nitride halides $M$N$X$ ($M$ = Ti, Zr, Hf; $X$ = Cl, Br, I). The parent compounds are band insulators with layered crystal structures and two-dimensional (2D) electronic states. 
In the layered nitrides, carrier density and interlayer distance can be controlled independently. 
Several unconventional superconducting 
properties beyond the conventional BCS theory have been reported. 

There are two kinds of layered polymorphs in metal nitride halides, $\alpha$- and $\beta$-forms with the FeOCl and the SmSI structures, respectively \cite{Yamanaka00, Yamanaka10}. The $\alpha$-structured polymorph has an orthogonal $M$N layer network separated by halogen layers as shown in Figure 1(a). The $\beta$-form consists of double honeycomb-like $M$N layers sandwiched by close-packed halogen layers as shown in Figure 1(b). Both polymorphs with nominal $M^{4+}$ 
($d^0$) 
configuration are band insulators with gaps larger than 2.5-4 eV, and therefore, there is neither magnetic order, such as in cuprates or pnictides, nor charge density wave order, as in BaBiO$_3$-related system. 
Electron doping is achieved by intercalation of alkali, alkaline-earth, or rare-earth metals between the Cl[$M_2$N$_2$]Cl layers, and interlayer distance can be controlled by cointercalation of solvent molecules. High-$T_c$ superconductivity with 
$T_c=25.5$ K was observed in the lithium and tetrahydrofuran (THF) cointercalated compound Li$_{0.48}$(THF)$_y$HfNCl (Figure 2) \cite{Yamanaka98}. The Zr homolog Li$_x$ZrNCl shows a $T_c=13$-15~K \cite{Yamanaka96, Taguchi06}. Metal nitride $M$N ($M$ = Ti, Zr, Hf) with the rock salt structure are well known 3D superconductors with $T_c$'s = 5.5, 10.7, and 8.8 K, respectively. Note that the electron-doped $\beta$-$M$N$X$ containing 2D $M$N layers separated by halogen and intervening dopant layers exhibit higher $T_c$'s than the corresponding 3D nitrides. 

In the following sections, 
we will comprehensively review the existing experimental and theoretical reports on layered nitride superconductors. The contents of this article are as follows: First, we review key experimental results in the normal and superconducting 
states of more extensively studied $\beta$-form materials. Next, the theories for the superconducting 
mechanism of the $\beta$-form materials are reviewed in Sec. 4. In Sec. 5, new superconducting 
$\alpha$-form materials 
\cite{Yamanaka09,Zhang12} are briefly reviewed. Finally, we summarize the normal and superconducting 
state properties, and compare them with other superconductors. 

\begin{figure}[t]
\begin{center}
\includegraphics[width=8cm]{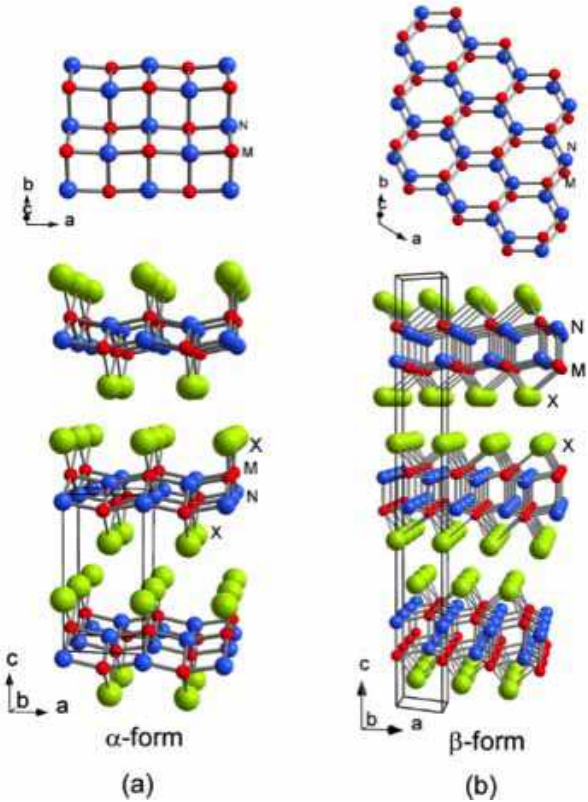}
\caption{
Schematic illustration of the crystal structures of (a) the $\alpha$- and (b) the $\beta$-forms of $M$N$X$ ($M$ = Ti, Zr, Hf; $X$ = Cl, Br, I): small red balls, $M$; blue balls, N; and large green balls, $X$. The lower part of the illustration shows the views along the $b$-axes, and the upper part shows the 2D nitride layers of each form \cite{Yamanaka10}.
}
\end{center}
\end{figure}

\begin{figure}[t]
\begin{center}
\includegraphics[width=7cm]{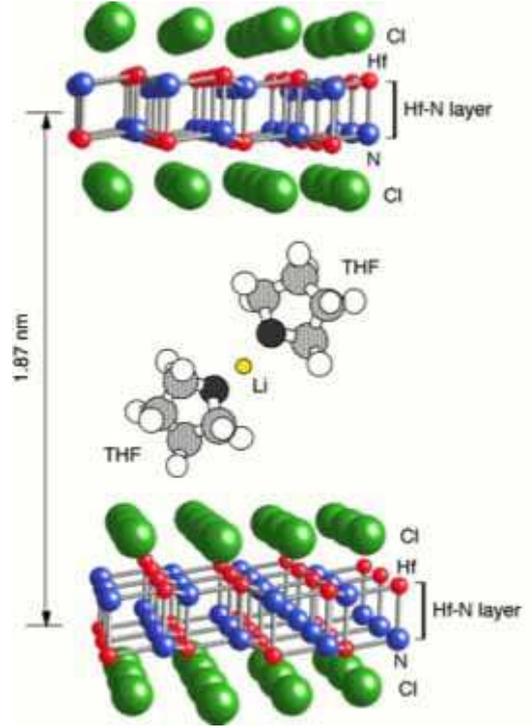}
\caption{
Schematic structural model for the Li and THF cointercalated $\beta$-HfNCl with $T_c=25.5$~K. The superconducting [HfN] layers are separated by chloride and the intervening cointercalation layers (dopants).
}
\end{center}
\end{figure}

\section{Normal state properties of the $\beta$-structured materials}

\subsection*{Band calculation}
Since the discovery of superconductivity in Li-doped $\beta$-HfNCl and $\beta$-ZrNCl \cite{Yamanaka96,Yamanaka98}, several groups have thus far performed band calculations \cite{Weht99, Hase99, Felser99, Sugimoto04, Heid05} to elucidate the basic electronic structure. All the calculation results indicate very similar band structures, and the band diagram calculated for Na$_{0.25}$HfNCl by Weht {\it {\it et al.}} \cite{Weht99} is shown in Figure 3. The conduction band shows a parabolic dispersion around K-point (and also at K$^\prime$-point) in the Brillouin zone for a hexagonal lattice, and the indirect gap is estimated to be approximately 2 eV. The conduction band consists of Hf 5$d_{xy}$ and 5$d_{x^2-y^2}$ orbitals (or Zr 4$d_{xy}$ and 4$d_{x^2-y^2}$ in case of ZrNCl) strongly hybridized with N 2$p$ orbitals while the valence band is mainly of N 2$p$ character. It is clear from 
Figure 3 
that a single band 
around the K-point 
is relevant to the low energy physics at low carrier density. The in-plane effective mass is estimated to be approximately 0.6 $m_0$ ($m_0$ denotes free electron mass). The conduction band has a highly 2D 
feature, 
as exemplified by flat dispersion along $\Gamma$A direction in Figure 3. Due to this two-dimensionality, Fermi surface of the electron-doped materials is two degenerate cylinders that locate around the K and K$^\prime$ point, 
and the density of states 
is nearly constant 
for the small carrier density regime.

\begin{figure}[t]
\begin{center}
\includegraphics[width=8cm]{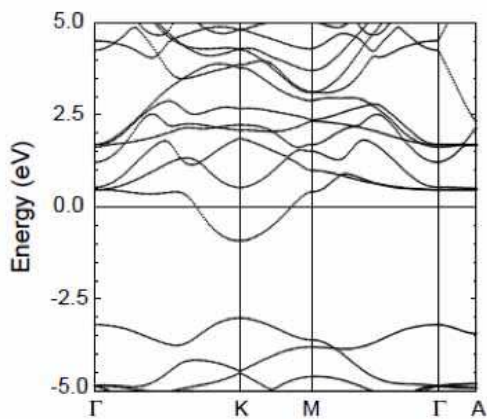}
\caption{
Band structure of Na$_{0.25}$HfNCl. $\Gamma=(0,0,0)$, K~$=(2/3,1/3,0)$, M~$=(1/2,0,0)$, A~$=(0,0,1/2)$ in units of the hexagonal reciprocal lattice vectors. Reproduced from \cite{Weht99}.
}
\end{center}
\end{figure}

\subsection*{X-ray absorption spectroscopy}
2D nature of the electronic structure has been confirmed by various experimental techniques. Figure 4 shows the intensity of X-ray absorption spectra at the N 1$s$ absorption edge for compressed samples of pristine HfNCl and Na-doped one, with $c$-axis almost perpendicular to the sample surface \cite{Yokoya01}. When the incident angle ($\theta$) is small, and hence electric-field vector is close to in-plane configuration, the intensity is stronger at the absorption edge for both compounds, demonstrating that the wave function at the conduction band bottom consists of the orbitals that are strongly hybridized with N $p_x$ and $p_y$ while hybridization with $p_z$ is small. 
This result clearly indicates the highly anisotropic electronic states, and is 
in good accord with the prediction of the band calculations \cite{Weht99, Hase99, Felser99, Sugimoto04, Heid05}.

\begin{figure}[t]
\begin{center}
\includegraphics[width=6cm]{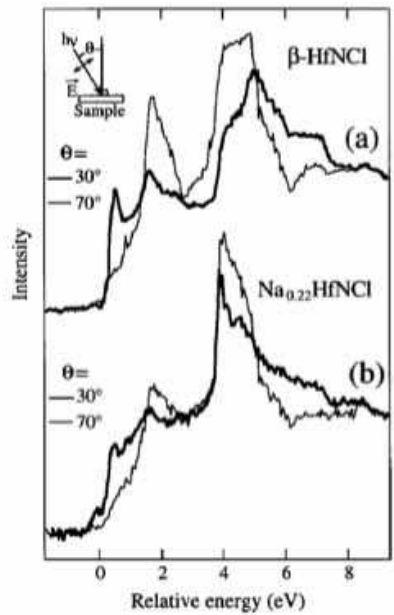}
\caption{
N 1$s$ X-ray absorption spectra of (a) $\beta$-HfNCl and (b) Na$_{0.22}$HfNCl in the total fluorescence mode. Measurement configuration is displayed as an inset, and $\theta$ denotes incident angle of the X-ray with respect to the sample normal. Reproduced from \cite{Yokoya01}
}
\end{center}
\end{figure}

\subsection*{Optical spectroscopy}
The overall electronic structure has been investigated by optical spectroscopy technique for Li$_x$ZrNCl system with different Li concentrations \cite{Takano08a,Takano11}. The reflectivity for the undoped compound shows a typical spectral shape for an insulator, namely, almost frequency-independent low reflectivity 
for below 3~eV, accompanied by spiky structures below 0.1 eV due to phonons, and a conspicuous peak around 3.7 eV due to interband transition. Upon Li doping, high reflectance band appears at low frequency region with a clear plasma edge. The square of the apparent plasma frequency is found to increase almost linearly with the carrier density, giving the effective mass of the conduction band of approximately 0.9$m_0$. This value is in excellent accord with the estimated value ($=0.89m_0$) based on specific heat result \cite{Taguchi05}. The detailed carrier-density dependence is well reproduced by a first-principles calculation \cite{Takano11}. A recent theory also reproduced the doping dependence of the plasma frequency as well as the dielectric constant value of $\varepsilon_\infty=5$ \cite{Botana14} as observed in the experiment. These results clearly indicate that the band calculations describe the electronic structure quite well, and also that the rigid band picture can be justified when electron-type charge carriers are doped.

\subsection*{Transport properties}
Systematic investigation on the doping dependence of resistivity was reported for Li$_x$ZrNCl system by using compressed pellet samples \cite{Taguchi06}. As shown in Figure 5(a), the undoped compound shows insulating temperature dependence. As the Li is intercalated and electrons are doped into the conduction band, the resistivity is progressively reduced. Li doping with $x>0.05$ 
induces superconductivity at low temperatures, nevertheless, the resistivity shows an upturn above $T_c$ for all the samples shown in Figure 5(a). The normal state conductivity values at 5 K 
and at 9 T 
is plotted as a function of $x$ in Figure 5(b). This figure clearly shows an insulator-metal transition at around $x=0.05$ in this system. 

As for the absolute value, genuine 
resistivity 
of a single-crystalline material would be much lower 
than those measured for the compressed pellets, where conduction electrons are subject to strong grain boundary scatterings. In fact, from systematic experiments for various samples with different doping concentrations in combination with phenomenological consideration, it was conjectured that the intrinsic value of the resistivity for a single crystal without any grain boundary scattering would be smaller than the observed value by almost two orders of magnitude \cite{Takano08b}. Also, intrinsic mobility values of carriers at low temperatures were estimated to be between 65 and 110 cm$^2$/Vs \cite{Takano08b} for a wide range of doping concentration $0.06<x<0.37$, which is in reasonable agreement with the value of 50 cm$^2$/Vs as deduced from an electric-double-layer-gating experiment for a single-crystalline thin flake at 220 K \cite{Ye10}. 

Anisotropy of the resistivity between $c$-axis and in-plane directions is estimated to be at least 100 from a measurement for ZrNCl$_{0.7}$ \cite{Tou05}. Since the measurement was performed for compressed pellet with slight amount of misoriented grains, the estimated value corresponds to the 
lowest 
limit of the resistivity anisotropy, clearly indicating the 2D nature of this material.

\begin{figure}[t]
\begin{center}
\includegraphics[width=9cm]{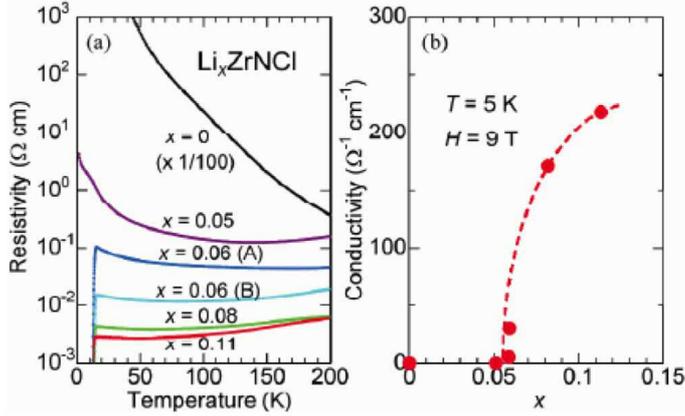}
\caption{
(a) Temperature dependence of resistivity for lightly doped Li$_x$ZrNCl. (b) Normal state conductivity at 5 K is plotted against doping concentration $x$. Superconductivity is suppressed by applying a magnetic field of 9 T. Reproduced from \cite{Taguchi06}.
}
\end{center}
\end{figure}

\subsection*{NMR and magnetic measurements}
2D nature of the electronic state was also confirmed by NMR and magnetic measurements for Li$_{0.48}$(THF)$_y$HfNCl at an early stage \cite{Tou00}. $^7$Li-NMR measurement revealed that Knight shift of Li in the normal state, and hence the Fermi level density of states 
at Li-site, is almost nothing, in good accord with the fact that the nature of the conduction band is mainly of planar 5$d_{xy}$ and 5$d_{x^2-y^2}$ character. DC-magnetization data around superconducting 
transition temperatures for various values of magnetic field were found to be well scaled in terms of lowest Landau level analysis, and it was concluded that this material is located at the border between a highly anisotropic 3D state and a 2D state \cite{Tou00}.   

Magnetic susceptibility measurement for a Hf-system \cite{Tou01a} clearly demonstrated  the small density of states 
at the Fermi level for their high-$T_c$ values, which is one of the most important characteristics of 
this class of 
superconductors. Figure 6 shows that the total magnetic susceptibility of the Li$_{0.48}$(THF)$_{0.3}$HfNCl is negative, indicating that the Pauli paramagnetic spin susceptibility is smaller than the sum of the contributions from ionic core and Landau diamagnetism. The total susceptibility value was carefully analyzed in terms of core diamagnetic susceptibility, Landau diamagnetism, and orbital susceptibility, and the spin susceptibility was determined to be $1.7\times10^{-5}$ emu/mol. This value is fairly small, and immediately implies the small density of states 
at the Fermi level, which was estimated to be 0.26 states/(eV spin f.u.) \cite{Tou01a}. By discussing this value in comparison with the results of band calculations and 2D free-electron model, it was also concluded that mass enhancement and exchange enhancement are both negligible. Furthermore, they conjectured that the electron-phonon interaction is weak, based on the consideration of superconducting 
energetics \cite{Tou01a}.

\subsection*{Specific heat}
Specific heat measurement for Li$_{0.12}$ZrNCl with $T_c$=12.7 K \cite{Taguchi05} also demonstrated the small density of states 
at Fermi level. The normal state Sommerfeld constant $\gamma_n$ was found to be $1.0\pm0.1$~mJ/mol K$^2$, which is fairly small for a superconductor with the $T_c$ around 10 K. For example, the 
$\gamma_n$ values for oxide and organic superconductors LiTi$_2$O$_4$ and $\kappa$-(BEDT-TTF)$_2$Cu[N(CN)$_2$]Br with similar $T_c$ values of 11.4 K and 11 K, respectively, have been reported to be 19.2 mJ/mol K$^2$ \cite{Sun04} and 22 mJ/mol K$^2$ \cite{Nakazawa00}, which are almost 20 times as large as that of Li$_{0.12}$ZrNCl. By using the density of states 
values (0.19-0.26 states/eV spin f.u.) predicted by band calculations, the upper limit of the electron-phonon coupling constant $\lambda$ is estimated to be 0.22. This value is also very small, and consistent with the predictions by the band calculation \cite{Heid05} as well as the conclusion deduced from magnetic susceptibility measurement for the 
Hf-system 
\cite{Tou01a}.

\subsection*{Raman scattering}
To probe the lattice dynamics and/or electron-phonon coupling, Raman scattering measurements have been performed \cite{Cros03, Kitora07}. Based on the measurements in various scattering configurations for HfNCl, ZrNCl, ZrNBr, and Na-doped HfNCl, Cros {\it et al.} have assigned the symmetry of the vibration modes for the observed six Raman active lines, and discussed the normal coordinates for each mode \cite{Cros03}. One of the Raman modes ($A_{1g}$ mode at 612~cm$^{-1}$ for HfNCl) was found to show a huge broadening upon the Na-intercalation, and this mode was suggested to be relevant to the superconductivity. Later, doping and temperature dependence of the phonon Raman scattering was investigated for a series of Li$_x$ZrNCl \cite{Kitora07}. Broadening of the phonon line width was analyzed in terms of disorder scattering, phonon-phonon scattering, and electron-phonon scattering, and the line width due to the respective contributions was separately determined. Among the observed modes, the highest frequency mode around 620 cm$^{-1}$ showed a systematic doping-variation of the thermal broadening, which was attributed to the electron-phonon scattering. As the doping concentration was increased, the thermal broadening of the line width became larger, and hence it was concluded that the electron-phonon interaction is larger for larger carrier density \cite{Kitora07}. This trend is opposite to the characteristic doping dependence of the superconducting 
transition temperature observed in the same system \cite{Taguchi06}.

\begin{figure}[t]
\begin{center}
\includegraphics[width=7cm]{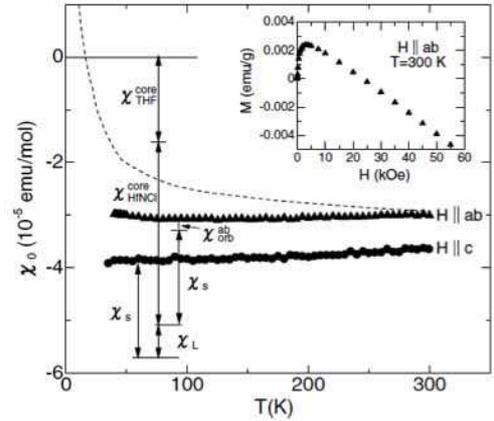}
\caption{
Temperature dependence of susceptibility of Li$_{0.48}$(THF)$_{0.3}$HfNCl for $H\parallel c$ and $H\parallel ab$. The dashed curve is the susceptibility of 
pristine 
$\beta$-HfNCl. Total susceptibility is negative, implying that Pauli paramagnetic spin susceptibility is very small. $\chi_\mathrm{core}$, $\chi_s$, $\chi_\mathrm{L}$, $\chi_\mathrm{orb}$ denote the core diamagnetic susceptibility, Pauli spin susceptibility, Landau diamagnetic susceptibility, and orbital susceptibility, respectively. Reproduced from \cite{Tou01a}.
}
\end{center}
\end{figure}

\section{Physical properties in the superconducting state of the $\beta$-structured materials}

\subsection*{Phase diagram and effect of interlayer distance on $T_c$}
Systematic studies with controlled doping and interlayer spacing by intercalation of metals and solvent (Solv) molecules have been performed for both $\beta$-ZrNCl and HfNCl. 
%
%
It has been reported that the stacking pattern of the layers Cl[$M_2$N$_2$]Cl is changed upon intercalation of metals, leading to the change of the polytype from the SmSI ($R\bar{3}m$) to the YOF type \cite{Shamoto04,Chen02} and to the $P\bar{3}m$ stacking \cite{Zheng11,Zhang13b,Zhang13c}. The changes depend on the kinds of the intercalated metals. It is reasonable to estimate that the polytype changes will hardly influence the $T_c$. 
%
%
Figure 7(a) shows the Li-doping $x$ dependence of $T_c$ in Li$_x$ZrNCl and Li$_x$(Solv)$_y$HfNCl \cite{Taguchi06, Takano08c}. 
With increasing $x$, superconductivity suddenly appears with finite $T_c$ at $x\sim0.05$ for Li$_x$ZrNCl and at 0.15 for Li$_x$(Solv)$_y$HfNCl, respectively, indicating insulator-superconductor transition. 
The aforementioned structural changes 
in polytype 
take place at 
Li 
concentrations lower than those for insulator-superconductor transitions in both materials. 
The observed 
insulator-superconductor transition without magnetic ordering is in marked contrast with  the Mott transition as observed in layered organic and fullerene superconductors \cite{Kanoda97,Ganin10}. One notable feature of this phase diagram is that the superconducting 
phase does not exhibit a dome-like shape which is observed in many superconductors, including SrTiO$_3$ \cite{Koonce67}, boron-doped diamond \cite{Ekimov04}, and cuprates \cite{Imada98}. The rapid increase of $T_c$ with reducing $x$ below $x = 0.12$ in Li$_x$ZrNCl is also a unique feature to this system 
(Figure 7(b)). 
The enhancement of $T_c$ with reducing doping cannot be reconciled with the weakening of the electron-phonon interaction as 
revealed 
by the Raman scattering experiments, within a simple framework of phonon-mediated superconductivity \cite{Kitora07}. The superconducting 
phase prevails in a wide range of doping, as shown by heavily doped Yb$_{0.31}$(NH$_3$)$_y$ZrNCl ($T_c\sim13$ K) \cite{Zhang13c} and Ca$_{0.4}$(NH$_3$)$_y$HfNCl ($T_c\sim24$ K) \cite{Zhang13b}. 

\begin{figure}[t]
\begin{center}
\includegraphics[width=9cm]{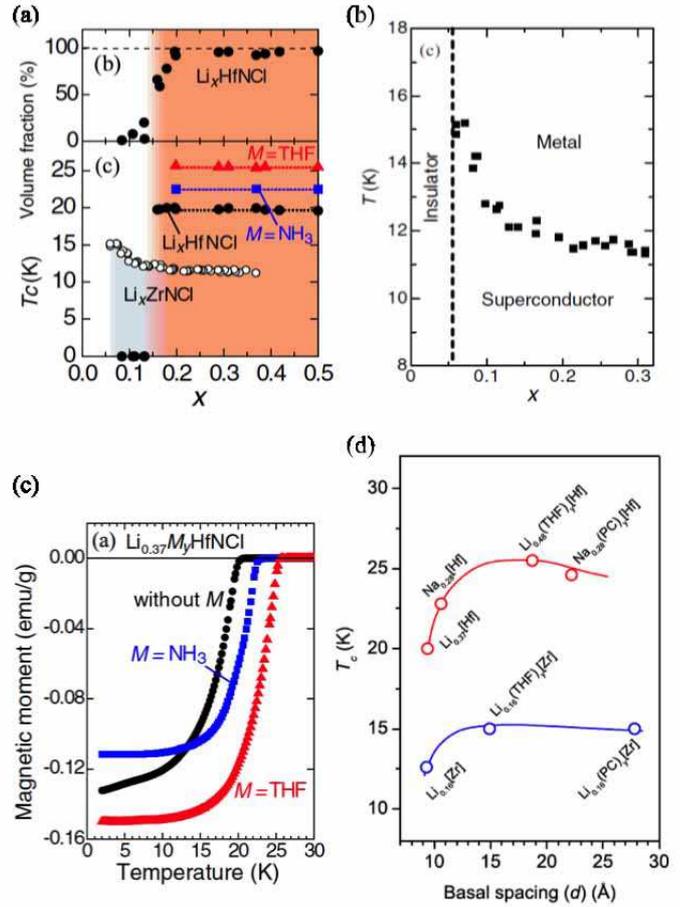}
\caption{
(a) $x$ dependence of the superconducting volume fraction for Li$_x$HfNCl and that of $T_c$ values for Li$_x$ZrNCl and Li$_x$(Solv)$_y$HfNCl \cite{Takano08c}. (b) Electronic phase diagram for Li-intercalated $\beta$-ZrNCl system. $x$ represents the Li concentration (Li$_x$ZrNCl). Reproduced from \cite{Taguchi06} (c) Temperature dependence of magnetization at 10 Oe for 
Li$_{0.37}$(Solv)$_y$HfNCl (Solv=NH$_3$ and THF). 
(d) $T_c$ vs the basal spacing ($d$) of the superconductors derived from $\beta$-ZrNCl ([Zr]) and HfNCl ([Hf]) \cite{Hotehama10}.
}
\end{center}
\end{figure}

Another important feature is that the enhancement of $T_c$ upon the cointercalation of neutral solvent molecules. Figure 7(c) shows the temperature-dependent magnetization curves for Li$_x$(Solv)$_y$HfNCl with different Solv molecules. $T_c$ increases from 20 K for Solv = none to 25.5 K for Solv = THF \cite{Takano08c,Yamanaka98}. $T_c$ values for molecule cointercalated ZrNCl and HfNCl are plotted as a function of the interlayer spacing $d$ in Figure 7(d) \cite{Hotehama10}. $T_c$ appears to attain a maximum value 
around 
$d\sim15$~\AA~in both ZrNCl and HfNCl systems. The highest $T_c=26.0$ K is obtained in Ca$_{0.11}$(THF)$_y$HfNCl \cite{Zhang13b}. With further increasing $d$, $T_c$ remains constant or decreases \cite{Hotehama10, Kasahara10}. The specific heat measurements of 
Li$_x$(Solv)$_y$ZrNCl 
have revealed that the enhancement of $T_c$ is ascribed to concomitant enhancement of the pairing interaction \cite{Kasahara10}. 
Based on these results, it is concluded 
that the pairing interaction strongly depends on the Fermi surface warping along the $k_z$ direction, implying the relevance of charge and/or spin fluctuations rather than phonons. 

\subsection*{Upper critical field}
Two-dimensionality of the superconducting 
state has been investigated by the upper critical field $H_{c2}$ measurements \cite{Tou00,Tou05}. 
The anisotropy ratio ($\gamma=H_{c2}^{ab}/H_{c2}^c\sim(dH_{c2}^{ab}/dT)/(dH_{c2}^c/dT)$), where $H_{c2}^{ab}$ and $H_{c2}^c$ are $H_{c2}$ for $H\perp c$ and $H\parallel c$, respectively, measures the anisotropy of superconducting states.
$\gamma$ = 14 and 3 were obtained for Li$_x$(THF)$_y$HfNCl and Li$_{0.17}$ZrNCl, respectively, from the 
slope of $H_{c2}(T)$ near $T_c$. However, the whole $H_{c2}(T)$ curves have not been determined, because of very high $H_{c2}$ at least above 20 T. Tou {\it et al.} have performed high-field magnetoresistance experiments as high as 27 T down to 0.5 K for Cl-deintercalated ZrNCl (ZrNCl$_{0.7}$, $T_c=13$ K) and determined the whole $H_{c2}(T)$ curves \cite{Tou05}. It is noteworthy that $H_{c2}(T)$ continues to increase linearly ($T_c-T$) as $T\rightarrow0$, in contrast with the conventional prediction, where $H_{c2}(T)$ saturates as $T\rightarrow0$. $\gamma=4.5$ obtained by this experiments provides a solid evidence for the quasi-2D nature of the superconducting 
state \cite{Tou05}.  

\begin{figure}[t]
\begin{center}
\includegraphics[width=6cm]{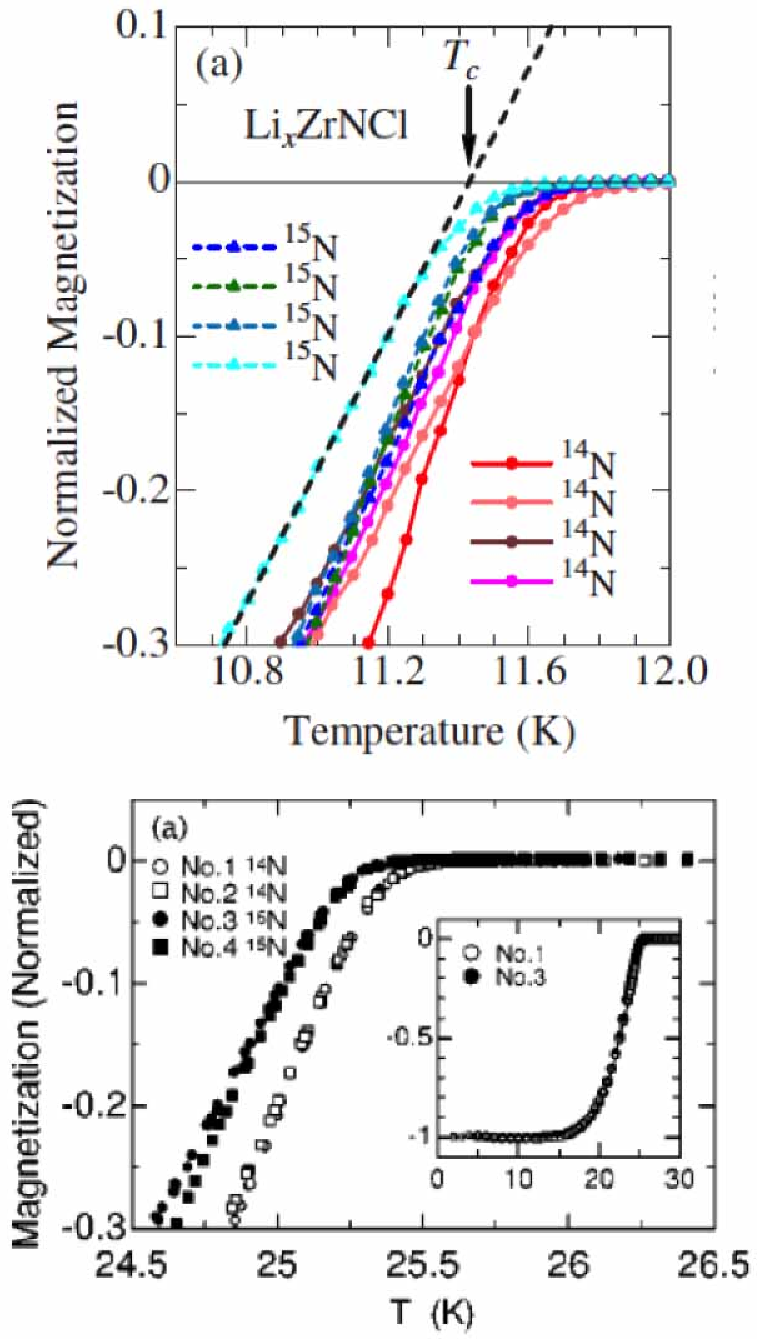}
\caption{
Temperature dependence of the magnetization around $T_c$ for $^{15}$N substituted (top) Li$_x$ZrNCl ($0.170\leqq x\leqq0.185$) \cite{Taguchi07} and (bottom) Li$_x$(THF)$_{0.30}$HfCNl \cite{Tou03}. 
}
\end{center}
\end{figure}

\subsection*{Isotope effect}
Isotope effect on $T_c$ is believed to provide key evidence whether electron-phonon interaction is responsible for superconductivity. In the standard BCS theory, 
$T_c\propto Z^{-\alpha}$ ($Z$ represents the atomic mass)  
and $\alpha=0.5$ are obtained. The N-isotope effect, replacing $^{14}$N to $^{15}$N, has been investigated for Li$_x$ZrNCl and Li$_x$(THF)$_y$HfNCl \cite{Tou03,Taguchi07}. Figures 8(a) and 8(b) show temperature dependences of the magnetization around $T_c$ for Li$_x$ZrNCl and Li$_x$(THF)$_y$HfNCl, respectively. The shift in $T_c$ ($\Delta T_c$) upon $^{15}$N substitution is $\Delta T_c=0.06$ K and 0.1 K for Li$_x$ZrNCl and Li$_x$(THF)$_y$HfNCl, respectively. The isotope shift coefficient has been found to be $\alpha\sim0.07$ for both Li$_x$ZrNCl and Li$_x$(THF)$_y$HfNCl, which is much smaller than $\alpha=0.5$. N-vibration phonon mode around 615 cm$^{-1}$ interacting most strongly with electronic system displays an softening by about 20 cm$^{-1}$ ($\Delta\omega/\omega=-3.3$ \%) upon $^{15}$N substitution, which is much stronger than the decrease in $T_c$ ($\Delta T_c/T_c\sim-0.5$ \%) \cite{Taguchi07}. These results may imply the relevance of other interactions than phonons to the pairing interaction. 

\begin{figure}[t]
\begin{center}
\includegraphics[width=9cm]{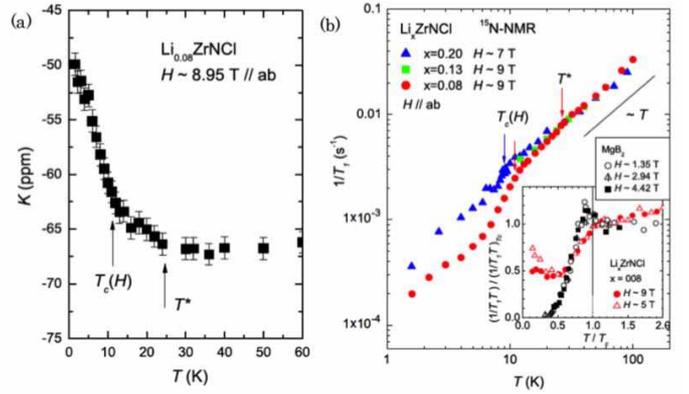}
\caption{
(a) Temperature dependence of the $^{15}$N NMR Knight shift, $K$, for Li$_{0.08}$ZrNCl. $K$ increases below $T_c(H)$ owing to spin-singlet paring, and a gradual variation of $K$ is observed below $T^\ast$ \cite{Kotegawa14}. (b) Temperature dependence of $^{15}$N-1/$T_1$ in Li$_x$ZrNCl for $x = 0.08$, 0.13, and 0.20. 1/$T_1$ is independent of $x$ in the normal state above $T^\ast$. A clear decrease below $T_c(H)$ appears in 1/$T_1$ without the coherence peak for $x = 0.08$. The inset shows a comparison between MgB$_2$ and Li$_x$ZrNCl \cite{Kotegawa14}. 
}
\end{center}
\end{figure}

\subsection*{NMR measurements}
NMR is a powerful probe to investigate the superconducting 
order parameter and magnetic fluctuations through the nuclear spin-lattice relaxation rate and the Knight shift. Spin-singlet superconductivity has been revealed by the Knight shift in Li$_x$ZrNCl and Li$_x$(THF)$_y$HfNCl \cite{Tou00,Tou02,Tou05,Kotegawa14}. Figure 9(a) shows the temperature dependence of the Knight shift in Li$_x$ZrNCl \cite{Kotegawa14}. The 
magnitude of the 
Knight shift 
decreases 
with decreasing temperature below $T_c$, indicating a decrease in spin susceptibility through a negative hyperfine coupling constant, 
providing  
evidence for the singlet pairing. 1/$T_1$ was firstly reported for Li$_x$(THF)$_y$HfNCl but suffers a large contribution from the vortex dynamics inherent in the 2D superconductor \cite{Tou02}, preventing the evaluation of intrinsic relaxation by quasiparticles. Recently, Kotegawa {\it et al.} have succeeded to obtain intrinsic 1/$T_1$ in Li$_x$ZrNCl where vortex dynamics is excluded because of the short interlayer distance \cite{Kotegawa14}. Figure 9(b) shows the temperature dependencies of $^{15}$N-1/$T_1$ in Li$_x$ZrNCl. 1/$T_1$ for $x = 0.08$ shows a clear decrease below $T_c$ without any signature of the coherence peak (Hebel-Slichter peak), which appears in the conventional BCS $s$-wave superconductors. $^{91}$Zr-1/$T_1$ in Li$_{0.08}$ZrNCl shows steeper decrease than $T^3$. An isotropic superconducting 
gap model with the complete exclusion of the coherence effect gives the superconducting 
gap ratio of $2\Delta_0/k_\mathrm{B}T_c = 4.5$, which is consistent with the strong-coupling superconductivity. 

\subsection*{Specific heat measurements}
The temperature and magnetic field dependences of the specific heat can probe low-energy 
elementary 
excitations. In the normal state, the Sommerfeld constant $\gamma_n$ is related to the density of states at Fermi energy. In superconductors having isotropic superconducting 
gap, the electronic specific heat divided by temperature, $\Delta C/T(T)$, shows an activated temperature dependence and $[\Delta C(H)-\Delta C(H=0)]/T=\gamma(H)\sim H$ at the lowest temperature. On the other hand, in superconductors having substantial anisotropy or nodal structure in the superconducting 
gap, $\gamma(H)$ will show a steep increase at low fields ($\sqrt{H}$ behavior in nodal $d$-wave superconductors). The specific heat of the nitride superconductors has been reported first for Li$_{0.12}$ZrNCl \cite{Taguchi05}, followed by Kasahara {\it et al.} for Li$_x$ZrNCl in a wide range of doping, $0.06\leqq x\leqq0.40$ \cite{Kasahara09}. Figures 10(a) and 10(b) show $\Delta C/T(T)$ for Li$_x$ZrNCl with $x = 0.10$ and 0.28, respectively. $\Delta C/T(T)$ at zero field can be described by the empirical theory with an assumption of isotropically gapped state, and the superconducting 
gap ratio $2\Delta_0/k_\mathrm{B}T_c$ of 4.14 for $x = 0.10$ and 2.9 for $x = 0.28$ has been obtained. The former clearly exceeds the BCS weak-coupling limit, $2\Delta_0/k_\mathrm{B}T_c=3.5$. 
As the doping concentration is reduced, 
superconducting 
coupling strength is concomitantly enhanced with $T_c$ while the density of states at the Fermi level is kept almost constant, indicating that the $T_c$ enhancement is due to the enhancement of the pairing interactions. As shown in Figure 10(c), $\gamma(H)$ shows $H$-linear behavior up to $H_{c2}$ for $x = 0.07$, close to the verge of the superconductor-insulator transition, which is consistent with isotropic gap. However, $\gamma(H)$ shows steep increase at low field and saturates much below $H_{c2}$ for $x > 0.10$, implying anisotropic superconducting 
gap. These results indicate that the anisotropy of the superconducting 
gap changes from almost isotropic to highly anisotropic upon 
increasing the doping level, 
and a possible pairing state with exotic symmetry other than $s$-wave is suggested.   

\begin{figure}[H]
\begin{center}
\includegraphics[width=7cm]{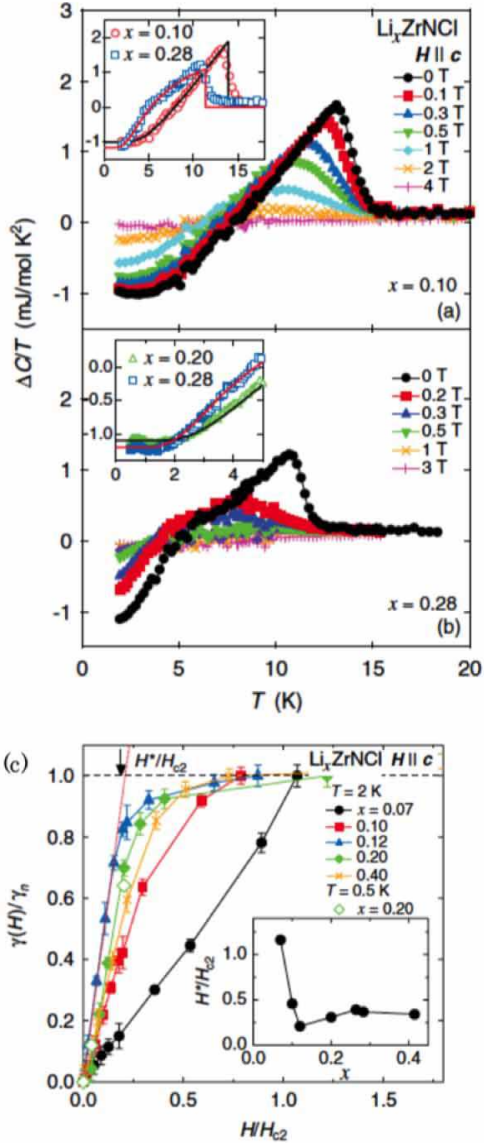}
\caption{
Electronic contribution of the specific heat $\Delta C/T$ as a function of $T$ in several magnetic fields 
for Li$_x$ZrNCl with (a) $x = 0.10$ and (b) 0.28. 
Inset: $T$ dependence of $\Delta C/T$ in zero field (open symbols) with the theoretical curves assuming the BCS relation (solid lines) is shown 
\cite{Kasahara09}. 
(c) Main panel: Normalized electronic specific heat coefficient $\gamma(H)/\gamma_n$ at 2 K as a function of magnetic field normalized by $x$-dependent $H_{c2}$. 
The dotted line is the linear extrapolation of initial slope for $x = 0.12$, and the $H^\ast$ is defined as the intersect of the dotted line and $\gamma(H)/\gamma_n = 1$ (dashed line). Inset: 
$H^\ast/H_{c2}$ 
as a function of $x$
\cite{Kasahara09}. 
}
\end{center}
\end{figure}
\noindent

\subsection*{$\mu$SR and Tunneling measurements}
Anisotropic superconducting 
gap was also suggested from the temperature and the magnetic field dependences of the muon depolarization rate $\sigma_s$, which is proportional to the superfluid density. For Li$_x$ZrNCl, Hiraishi {\it et al.} have found that the anisotropy becomes stronger with increasing $x$ \cite{Hiraishi10}, which is consistent with the specific heat measurements. Strong-coupling superconductivity was also confirmed by $\mu$SR experiments for both Li$_x$ZrNCl and Li$_x$(THF)$_y$HfNCl \cite{Hiraishi10,Ito04}. Ito {\it et al.} showed that the 2D superconducting 
carrier density $n_s^{2D}$ divided by effective mass $m^\ast$, $n_s^\mathrm{2D}/m^\ast$, rather than the 3D counter part, $n_s^\mathrm{3D}/m^\ast$, is the 
dominant factor for determining the high $T_c$ in the nitride superconductors, and suggested similarity to overdoped cuprates \cite{Ito04}, implying 2D nature and unconventional superconductivity in the nitride superconductors. 

Direct observation of the superconducting 
gap has been reported using break junction technique and scanning tunneling spectroscopy \cite{Takasaki05, Ekino13}. The superconducting 
gap ratio $2\Delta_0/k_\mathrm{B}T_c$ reaches 10 for ZrNCl$_{0.7}$, indicating strong coupling superconductivity. 

\subsection*{Pressure effect}
Pressure ($P$) dependence of $T_c$ has been reported in Na$_{0.3}$HfNCl, ZrNCl$_{0.7}$, and Li$_{0.5}$(THF)$_y$HfNCl \cite{Shamoto00, Taguchi04} and it has been found that $(1/T_c)(dT_c/dP)\sim-$(2.5-7)$\times10^{-3}$/GPa, which is one or two orders of magnitude smaller than that of MgB$_2$ and fullerene superconductors. $P$ dependence of the lattice constant and phonon frequency was also investigated \cite{Taguchi04}. The analysis in terms of McMillan's theory pointed out that low frequency phonons below 100 cm$^{-1}$ is responsible for the superconducting 
paring, which is in marked contrast with the theories and thermodynamic experiments \cite{Taguchi04}. 
This result should rather be regarded as indicating that McMillan's theory cannot be applicable to this superconductor. 

\begin{figure}[t]
\begin{center}
\includegraphics[width=9cm]{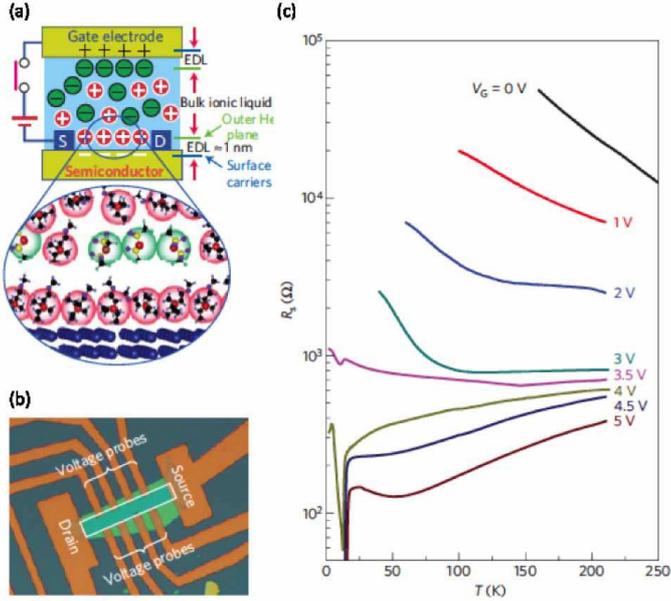}
\caption{
(a) (top) Schematic of charge accumulation by an EDL formed at interface between ionic liquid and semiconductor. Cations (red circles) and anions (green circles) can be electrostatically accumulated onto the channel surface by gate bias. (bottom) A magnified view of the semiconductor/ionic liquid interface and schematic models of molecules of cation and anion are also shown. (b) A ZrNCl thin flake patterned into a Hall bar configuration. The channel of the transistor is indicated by the white rectangle. (c) Temperature dependence of channel sheet resistance $R_s$ at different gate biases $V_G$ \cite{Ye10}. 
}
\end{center}
\end{figure}

\subsection*{Electric-field-induced superconductivity}
In the above, we showed the physical properties of the bulk superconductivity induced by chemical doping. An alternative way of carrier doping is electrostatic charge accumulation using a field-effect-transistor (FET) device. Recently, electric double layer transistor (EDLT) opens a new route for creating superconducting 
phases through the electric-field-effect from insulating materials \cite{Ueno14}. EDLT is a variation of FETs (Figure 11(a)) where an ionic conductor, such as ionic liquid, is used as a gate dielectric. Under an applied gate voltage, ions and charge carriers accumulated at the interface form an electronic double layer (EDL) that acts as an interface capacitor with nanoscale thickness. Electrostatic charge accumulation in EDLTs rises to a new benchmark for a 2D carrier density $n_\mathrm{2D}$ up to $\sim10^{15}$ cm$^{-2}$, which is larger than the critical carrier density for the nitride superconductors, $\sim10^{14}$ cm$^{-2}$. 

Ye {\it et al.} have fabricated an EDLT device using ZrNCl as a channel material (Figure~11(b)) \cite{Ye10}. For device fabrication, mechanical micro-cleavage techniques have been applied as in graphene, to make atomically flat surfaces. With applying gate voltage $V_G$, the sheet resistance $R_s$ was dramatically reduced, followed by an occurrence of the insulator-metal transition. Superconducting 
transitions with zero resistance were observed for $V_G$ of 4.5 V and 5 V 
(Figure 11(c)). 
Charge accumulation by EDL does not introduce disorder, providing an opportunity to investigate intrinsic physical properties of the materials.

\section{Theory of superconducting mechanism of the $\beta$-structured materials}

\subsection*{Possibility of phonon-mediated pairing}

Theoretically, there are a number of puzzles to be resolved for the layered nitride superconductors. The theory needs to explain various curious observations, some of which may seem contradicting:  the relatively high $T_c$ despite the low density of states \cite{Tou01a,Taguchi05,Weht99}, fully gapped, but still anisotropic superconducting 
state \cite{Taguchi05, Kasahara09, Hiraishi10}, and also the absence of the coherence peak in the NMR experiment \cite{Kotegawa14} as well as the small isotope effect \cite{Tou03, Taguchi07}. Some early theoretical studies used first-principles band calculation to investigate the possibility of phonon-mediated pairing \cite{Weht99, Heid05}. Using the phonon dispersion and solving the gap equation numerically, only $T_c<5$ K was obtained for a reasonable choice of the Coulomb repulsion pseudopotential $\mu^\ast$. When $\mu^\ast$ is set to be $\sim0$, a realistic $T_c$ can be obtained, but then the nitrogen isotope effect is estimated to be too large compared to the experiment \cite{Heid05}. This is a consequence of the low density of states \cite{Tou01a,Taguchi05} and the moderate electron-phonon coupling as observed experimentally. 

Recently, there have been more advanced theoretical studies on the possibility of phonon-mediated pairing. Yin {\it et al.} used a hybrid functional \cite{HSE} instead of LDA/GGA functionals in the first-principles calculation, and estimated the electron-phonon coupling \cite{Yin13}. This takes account of the dynamic correlation effects that are not appropriately treated in LDA/GGA. The estimated electron-phonon coupling $\lambda$ is larger than those obtained within LDA/GGA and is even close to unity in HfNCl. Using the modified McMillan's formula and assuming $\mu^\ast=0.1$, they find a $T_c$ quite close to the experimental values. 

\begin{figure}[t]
\begin{center}
\includegraphics[width=7cm]{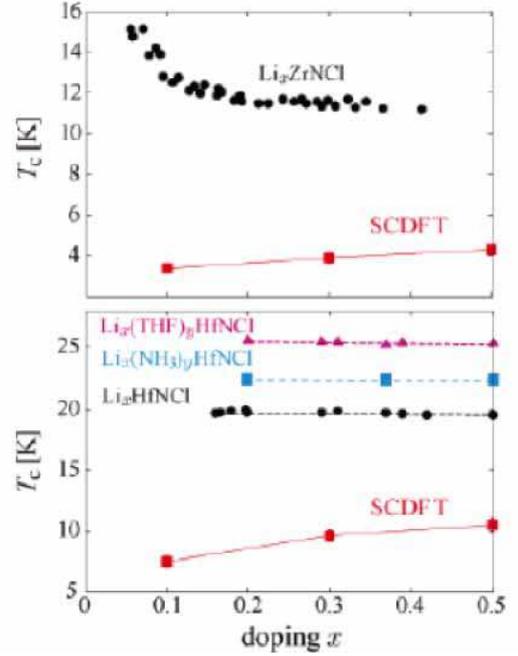}
\caption{
$T_c$ calculated using SCDFT \cite{Akashi12} compared with the experimental results.
}
\end{center}
\end{figure}

On the other hand, Akashi {\it et al.} used density functional theory for superconductors (SCDFT) \cite{Akashi12}. This serves as a parameter-free method to calculate $T_c$, where electron-phonon and electron-electron interactions are treated nonempirically \cite{Luders05}, and hence adjustable parameters like $\mu^\ast$ are not present. This method has been applied to various phonon-mediated superconductors such as simple metals \cite{Marques05}, MgB$_2$ \cite{Floris05}, $etc$. In Figure 12, we show the $T_c$ obtained by this formalism for the layered nitride superconductors. It is seen that the theoretical $T_c$ is much lower than the experimental observations. This result reinforces the previous calculations based on the McMillan's formula that electron-phonon interaction is not strong enough to account for the high $T_c$. This study, however, uses the GGA functional, so that there remains a possibility that using the hybrid functional as in ref. \cite{Yin13} may enhance the electron-phonon coupling and $T_c$. In any case, regarding the electron-phonon scenario, there are some experimental observations that have to be explained aside from $T_c$; experimental evaluation of a small electron-phonon coupling ($\lambda\sim0.2$) \cite{Taguchi05}, the very small isotope effect \cite{Tou03,Taguchi07}, absence of the coherence peak in the NMR experiments \cite{Kotegawa14}, and the anisotropy of the superconducting gap \cite{Taguchi05,Kasahara09,Hiraishi10}. Also, the increase of $T_c$ upon approaching the insulating state in Li$_x$ZrNCl \cite{Taguchi06, Kasahara09} is puzzling because there seems to be no reason for the electron-phonon interaction and/or the density of states to become large in the small carrier concentration regime. We will come back to this last point in the end of this section.

\subsection*{Plasmon-assisted pairing mechanism}

Owing to the above mentioned theoretical as well as experimental indications that the pairing of the electron-doped $\beta$-$M$NCl may not be mediated purely by phonons, there have been several proposals for unconventional pairing mechanisms. Bill {\it et al.} focused on the role played by acoustic plasmons, that is, the low energy plasmons that have linear dispersion near $k=0$ \cite{Bill03}. Originally, Takada pointed out the possibility of pairing mediated by acoustic plasmons in layered systems \cite{Takada88}. In ref. \cite{Bill03}, the authors constructed a general theory for layered systems, where the phonon-mediated pairing interaction is strong enough to have superconductivity at $T=0$, and considered the possibility of $T_c$ enhancement due to acoustic plasmons. Using the electron-phonon coupling $\lambda$ deduced from the 
specific heat results 
($\lambda\sim0.25$), they obtained a $T_c$ consistent with the experiments, showing that the acoustic plasmons can strongly enhance the pairing. Since this approach contains some parameters, it is interesting to investigate this possibility in a more first-principles manner. In fact, quite recently, Botana and Pickett have performed a first-principles study on the plasmon in Li$_x$ZrNCl \cite{Botana14}. 
$T_c$ itself was not calculated, but it has been pointed out that the doping level dependence of the reflectivity and the energy loss function seems to be uncorrelated with the nearly doping independent $T_c$ observed experimentally.

\subsection*{Spin-fluctuation scenario}

Another proposal of unconventional pairing is based on Hubbard-type models with short range repulsive electron-electron interactions. Kuroki \cite{Kuroki10} introduced a two-band tight-binding model that reproduces the band structure near the Fermi level obtained in the first-principles calculations \cite{Weht99, Heid05, Sugimoto04, Hase99}. In real space, this corresponds to considering both the $M$ (Zr or Hf) and the nitrogen sites on the honeycomb lattice. Originally, the introduction of such a model was based on the idea that the Hubbard model on the honeycomb lattice can give $d$-wave superconductivity with relatively high $T_c$ \cite{Kuroki09}. In fact, applying the spin-fluctuation-mediated pairing (fluctuation-exchange approximation \cite{Bickers89}) theory to the Hubbard model with $U=6t$ on the canonical honeycomb lattice gives $d$-wave superconductivity with a $T_c$ of $0.005t$ for a band filling near half-filling. Here $t$ is the nearest neighbor hopping integral, and if this is taken to be $\sim1$~eV, then $T_c\sim50$ K.The honeycomb lattice is a bipartite lattice, namely, it  can be decomposed into two sublattices A and B. When an energy level offset $\Delta$ is introduced between the two sublattices, a gap opens at the center of the band, and $T_c$ of the spin-fluctuation mediated pairing goes down. Interestingly, the decreasing rate of $T_c$ with the increase of $\Delta$ in the case of the honeycomb lattice is rather small compared to that obtained by the same model for the square lattice \cite{Kuroki10,Kuroki09}. 

\begin{figure}[t]
\begin{center}
\includegraphics[width=6.5cm]{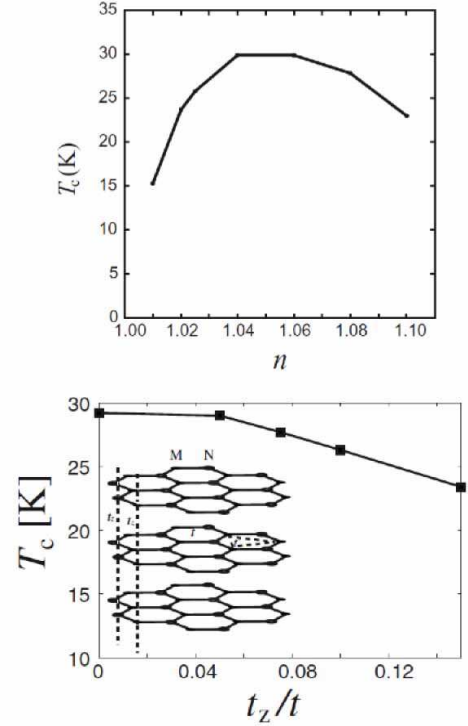}
\caption{
(top) $T_c$ of the two-band model of $\beta$-$M$NCl plotted as a function of $n$, the electron band filling \cite{Kuroki09}. (bottom) $T_c$ of the two-band model with finite interlayer hopping $t_z$ \cite{Kuroki10}.
}
\end{center}
\end{figure}

In Ref. \cite{Kuroki10,Kuroki09}, this theory was applied to the two-band model that has a more realistic band structure. $T_c$ of the model is shown in Figure13 (top) as a function of the band filling $n$. Here $n-1$ corresponds to the electron doping rate. In Figure 13 (bottom), we show the $T_c$ calculation result of the 3D model that considers the interlayer hopping $t_z$. It is seen that the introduction of the three dimensionality degrades superconductivity.

\subsection*{$d+id$ pairing}

As for the paring symmetry within the spin-fluctuation scenario, 
the two-band model on the honeycomb lattice gives $d$-wave pairing, which is two-fold degenerate due to the symmetry of the lattice. Hence, one of the possible pairing states below $T_c$ is the fully gapped $d+id$ state, where the gap function is complex and its real and imaginary parts have $d_{x^2-y^2}$ and $d_{xy}$ symmetries, respectively. This state has been proposed as the most stable pairing state for Hubbard-type models on the triangular lattice \cite{Baskaran03, Kumar03, Ogata03, Zhang06}, or on the honeycomb lattice as models for graphene \cite{Chubukov12,DHLee12,Thomale12}. In Figure 14, we show the absolute value of the $d+id$ superconducting 
gap constructed from the $d$-wave gap obtained within the fluctuation-exchange approximation for the two-band model \cite{Kuroki10}. When the doping rate is small, the gap is nearly isotropic on the Fermi surface, but as the doping rate is increased, the gap becomes anisotropic.

\begin{figure}[t]
\begin{center}
\includegraphics[width=7cm]{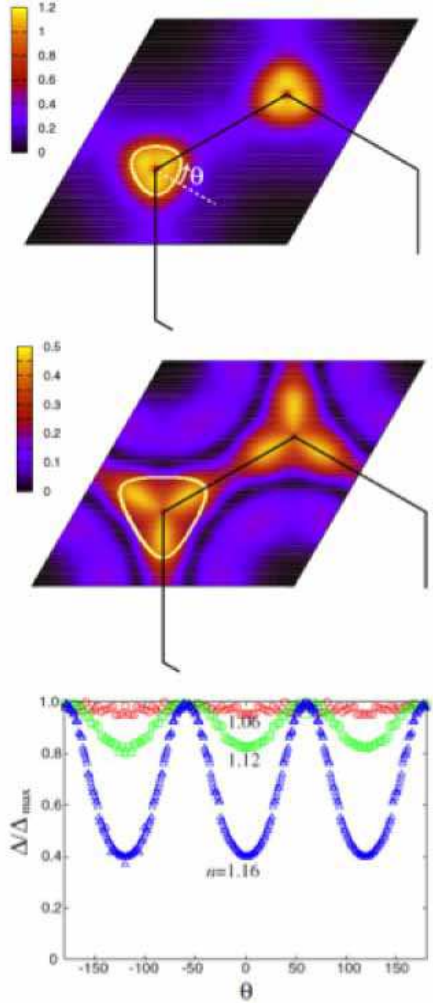}
\caption{
Contour plots of the absolute value of the $d+id$ gap obtained for the two-band model with $n=1.06$ (top) and $n=1.16$ (middle). The solid line closed around the Brillouin zone edge represents the Fermi surface. Lower panel: the normalized absolute value of the $d+id$ gap along the Fermi surface. The angle $\theta$ is defined in the top panel \cite{Kuroki10}.
}
\end{center}
\end{figure}

The spin-fluctuation scenario enables us to explain some of the seemingly contradicting experimental observations. Most importantly, the absolute value of the $d+id$ gap is finite on the entire Fermi surface, but the gap function itself integrated over the Fermi surface is exactly zero, so that the coherence peak should be absent in 1/$T_1$, in agreement with the NMR experiment \cite{Kotegawa14}. The enhancement of the gap anisotropy with increasing doping \cite{Taguchi05,Kasahara09,Hiraishi10} as well as the enhancement of $T_c$ with increasing two dimensionality in $\beta$-$M$NCl \cite{Takano08c} is also consistent. Nonetheless, this theory also confronts difficulties. First, although the uniform magnetic susceptibility was found to be enhanced in Li$_x$ZrNCl upon decreasing the doping rate \cite{Kasahara09}, the NMR experiment shows no enhancement of 1/$T_1T$ at low temperatures for small doping concentrations, indicating the absence of low energy spin fluctuations \cite{Kotegawa14}. Secondly, a variational Monte Carlo study has been performed for the similar two-band model \cite{Watanabe13}, where a nodal $d$-wave state is found to be more stable than the $d+id$ state for finite $\Delta$ (A-B sublattice energy offset). If that is the case, the gap will be nodal on the Fermi surface and inconsistent with the experiment. Thirdly, this mechanism relies on the peculiar features of the honeycomb lattice, so that it cannot be applied to the $\alpha$-form materials (see Section 5), automatically implying that the pairing mechanisms of the two phases are different. And finally, it is again difficult to theoretically reproduce the $T_c$ enhancement in the lightly doped regime \cite{Taguchi06}.

As for the NMR, since 1/$T_1T$ measures Im $\chi(q,\omega)/\omega$ in the $\omega\rightarrow0$ limit, there is a possibility that the  finite energy spin fluctuations that are not detected in the NMR may be responsible for superconductivity. On the other hand, a more straightforward interpretation of course is that the spin fluctuation does not play a major role in the pairing. Even if that is the case, $d+id$ pairing is still a tempting possibility as an explanation for seemingly contradicting observations. Then, it would be interesting to look into the possibility of such pairing owing to fluctuations other than the spin fluctuation.  Such a possibility could be investigated in more realistic models which consider not only on-site but also off-site electron-electron interactions as well as more realistic band structures with more than just two bands. 

\subsection*{Possible relevance of the Anderson localization}

Regarding the $T_c$ enhancement in the lightly doped regime, apart from the microscopic pairing mechanism, the system approaching the Anderson localization may be playing some role in the enhancement of $T_c$, as suggested by several authors \cite{Feigelman07, Yanase09, Burmistrov12, Kotegawa14}. The Anderson localization scenario may also account for the absence of the coherence peak \cite{Kotegawa14}. Nonetheless, the relatively high $T_c$ as well as the absence of the coherence peak holds even in the largely doped regime, where the localization effect is expected to be weak, so once again puzzles remain. It is worth pointing out that a similar phase diagram, where the insulating and superconducting 
phases sit next to each other has recently been observed in a completely different (but with layered structure) superconductor, Sr$_{1-x}$La$_x$FBiS$_2$ \cite{Sakai14}.

\section{Superconductivity of the electron doped $\alpha$-structured TiNCl by intercalation}

Recently the $\alpha$-structured layered TiNCl has been electron-doped by the intercalation of alkali metals as well as organic Lewis bases, which turned out to become superconductors with relatively high $T_c$'s \cite{Yamanaka09, Yamanaka12, Zhang12}. As mentioned in the above section, the structures of the two kinds of $M$N$X$ ($\alpha$- and $\beta$-forms) are very different within the layers, and accordingly they have different types of band structures. It will be interesting to compare their superconductivities. The $\beta$-structured compounds have disconnected cylindrical Fermi surfaces favorable for nesting \cite{Kuroki07, Takano08c, Felser99}, while the electron-doped $\alpha$-structured TiNCl has a simple oval shaped Fermi surface around 
only 
the $\Gamma$ point \cite{Yin11}, suggesting that a similar possibility of the nesting is excluded. 

\begin{figure}[t]
\begin{center}
\includegraphics[width=9cm]{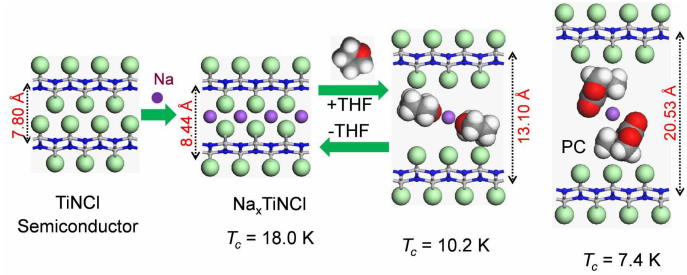}
\caption{
Schematic illustration showing the expansion of the basal spacing ($d$) of $\alpha$-TiNCl upon intercalation of Na and cointercalation of solvent molecules.
}
\end{center}
\end{figure}

\begin{figure}[t]
\begin{center}
\includegraphics[width=8cm]{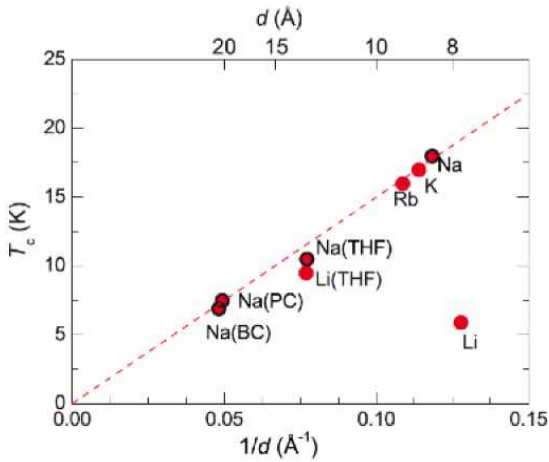}
\caption{
$T_c$ vs $1/d$ for $A_x$TiNCl ($A$ = Li, Na, K, Rb) superconductors with and without cointercalation \cite{Zhang12}.
}
\end{center}
\end{figure}

\subsection*{Alkali metal intercalation in TiNCl}
TiNCl was subjected to reaction with alkali metal naphthalene solutions in THF \cite{Zhang12}. The as-prepared compound is cointercalated with THF. The THF molecules can be removed by evacuation, or replaced with different kinds of solvent molecules such as PC (propylene carbonate) 
and BC (butylene carbonate), 
keeping the doped metal concentration unchanged. As shown in Figure 15, the basal spacing ($d$) of the compound varies depending on the cointercalation conditions, and the $T_c$ decreases with the increasing $d$. 
Figure 16 shows the $T_c$'s as a function of $1/d$ for the alkali metal intercalation compounds with and without cointercalation \cite{Zhang12}. The data fit on a linear line passing through the origin, suggesting the importance of the Coulomb interlayer coupling in the pairing mechanism in this system. The $T_c$'s of $A_x$TiNCl ($A$ = Na, K, Rb) also fit on the same line. It had been expected that Li$_x$TiNCl with the smallest basal spacing $d$ (7.8 \AA) should have shown the highest $T_c$ in this system. However, the $T_c$ was found to be as low as 6.0 K. This discrepancy could be attributed to the different types of charge distribution; Li ions are so small in size to penetrate into chlorine layers, forming double LiCl layers between [TiN] layers. In the THF cointercalated compound, Li$_{0.13}$(THF)$_y$TiNCl ($d = 13.1$ \AA), Li ions can be coordinated with THF molecules between chlorine layers like Na ions in Na$_x$(THF)$_y$TiNCl of Figure 15, and the $T_c$ fits again on the linear line as shown in Figure 16. Another systematic study on the electron-doped $\alpha$-structured TiNBr has also demonstrated a very similar linear relation on the $T_c$ vs $1/d$ \cite{Zhang13a}. 

The Rietveld analysis of the X-ray powder diffraction data revealed that the TiNCl crystalline layers are mutually shifted to accommodate metal atoms with different sizes between the chloride layers 
\cite{Yamanaka09,Zhang12}. 
However, it is reasonable to assume that the polytype changes will hardly influence the electrical properties of layer structured compounds. Note that the $d$ dependence of $T_c$ in the $\beta$-structured compounds described in Section 3 is opposite to that of the $\alpha$-structured compounds. The opposite dependence is explained in terms of topologically more favorable nesting in the $\beta$-structured compounds upon the expansion of the spacing \cite{Takano08c}.

\begin{figure}[t]
\begin{center}
\includegraphics[width=6cm]{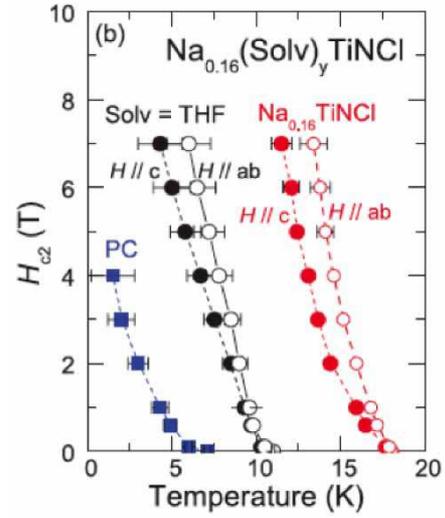}
\caption{
$H_{c2}$-$T$ phase diagram of Na-intercalated compounds with and without cointercalation \cite{Zhang12}.
}
\end{center}
\end{figure}

\begin{table*}[t]
\begin{center}
\caption{Characteristic superconducting parameters of the $\alpha$-form and the $\beta$-form nitride superconductors \cite{Zhang13a}}
\begin{tabular}{ccccccccc}
\hline
Compound & $T_c$ (K) & $d$ (\AA) & $\xi_{ab}$ (\AA) & $\xi_c$ (\AA) & 
$\gamma$ & Reference\\
\hline
$\alpha$-K$_{0.21}$TiNBr & 17.2 & 9.5 & 53 & 41 & 
1.3 & \cite{Zhang13a}\\
$\alpha$-Na$_{0.16}$TiNCl & 18.1 & 8.4 & 33 & 28 & 
1.2 & \cite{Zhang12}\\
$\alpha$-Na$_{0.16}$(THF)$_y$TiNCl & 10.2 & 13.1 & 55 & 35 & 
1.5 & \cite{Zhang12}\\
$\beta$-ZrNCl$_{0.7}$ & 13 & 9.8 & 71 & 16 & 
4.5 & \cite{Tou05}\\
$\beta$-Li$_{0.48}$(THF)$_y$HfNCl & 25.5 & 18.7 & 60 & 16 & 
3.7 & \cite{Tou01b}\\
$\beta$-Eu$_{0.08}$(NH$_3$)$_y$HfNCl & 24.3 & 11.9 & 61 & 15 & 
4.1 & \cite{Zhang13c}\\
$\beta$-Ca$_{0.11}$(THF)$_y$HfNCl & 26.0 & 15.0 & 47 & 12 & 
4.1 & \cite{Zhang13b} \\
\hline
\end{tabular}
\end{center}
\end{table*}


\subsection*{Anisotropic superconducting properties}
 Based on the anisotropic upper critical fields ($H_{c2}$'s) of the layered superconductors Na$_{0.16}$(Solv)$_y$TiNCl measured on the highly oriented pellet samples (Figure 17), the 
anisotropy 
parameter $\gamma = (dH_{c2}^{ab}/dT)/(dH_{c2}^c/ dT) = \xi_{ab}/\xi_c$  (ratio of the coherence lengths in the $ab$ plane and along the $c$ axis) was determined to be $\gamma$ = 1.5 and 1.2 for Na$_x$(THF)$_y$TiNCl and Na$_x$TiNCl, respectively \cite{Zhang12}. These values suggest that the TiNCl superconductors exhibit rather isotropic or 3D character, and the expansion of the basal spacing from 8.44 \AA~
to 13.10 \AA~upon cointercalation of THF has little effect on the anisotropy. The superconductor derived from TiNBr also shows a similar $\gamma$ value \cite{Zhang13a}. The characteristic superconducting 
 parameters of the $\alpha$- and $\beta$-structured layered nitride superconductors are compared in Table 1. The small anisotropic parameter $\gamma$ appears to be one of the characteristic features of the $\alpha$-structured layered superconductors. This is 
in marked 
contrast to the finding that the $\beta$-structured nitrides show the anisotropic parameter $\gamma$ as large as 3.7-4.5 \cite{Zhang13b,Zhang13c,Tou01b,Tou05}. 
$\xi_c$
of $\beta$-Li$_{0.48}$(THF)$_y$HfNCl is about 16 \AA, comparable with the basal spacing 17.8 \AA, $i.e.$, the separation of the superconducting 
layers; the superconducting 
$\beta$-structured layers may be weakly Josephson coupled. On the other hand, in the electron-doped TiNBr and TiNCl, the $\xi_c$'s are more than three times larger than the basal spacing, implying that the $\alpha$-structured nitride layers are more strongly coupled. 

It is interesting to note that $\beta$-ZrNCl$_{0.7}$ (Table 1), which is electron-doped by a partial deintercalation of chlorine atoms from the interlayer space, has an anisotropic parameter as large as 4.5 
\cite{Tou05}. 
In $\beta$-ZrNCl$_{0.7}$ with $d = 9.8$ \AA, the nitride layers should be directly coupled without intervening alkali atoms. Nevertheless, the anisotropic parameter is comparable to, or even larger than, that of the cointercalated compound $\beta$-Li$_{0.48}$(THF)$_y$HfNCl. It is evident that the interlayer separation is not a decisive parameter for the anisotropy on $H_{c2}$. For more discussion on the superconducting 
mechanisms and the anisotropy of the two different kinds of layered nitrides, a theoretical study including the electronic 
band structures is required. Similar discussions should be done on the anisotropy of layered cuprates and iron pnictides in comparison with the layered nitride superconductors \cite{Zhang13a}. 

\begin{figure}[t]
\begin{center}
\includegraphics[width=7cm]{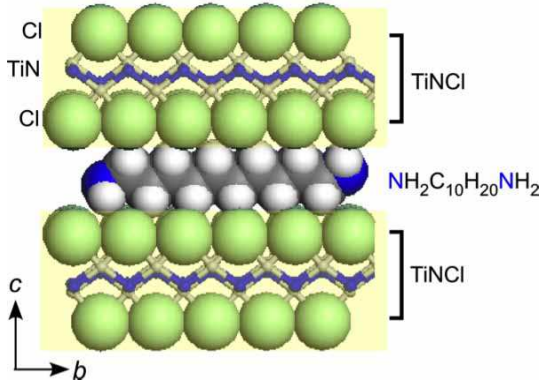}
\caption{
A schematic illustration of the arrangement of decamethylene diamine molecules between TiNCl layers with the molecular axis aligned along the $b$ axis of the crystal. The intercalation compound (NH$_2$C$_{10}$H$_{20}$NH$_2$)$_{0.12}$TiNCl becomes a superconductor with $T_c=17.1$~K \cite{Yamanaka12}. 
}
\end{center}
\end{figure}

\subsection*{Intercalation compounds of TiNCl with neutral amines}

TiNCl can form intercalation compounds with organic molecules alone without the association of metal intercalation. The intercalation compound with pyridine (Py, C$_5$H$_5$N), Py$_{0.25}$TiNCl expands the basal spacing to 13.5 \AA~with the molecular plane of Py oriented perpendicular to the layers, and becomes a superconductor with $T_c = 8.6$ K \cite{Yamanaka09}. Although the doping mechanism is not yet clear in the intercalation compound with Py, it would be reasonable to expect that the lone pair electrons of nitrogen atoms in Py may act as electron donors to the TiNCl layers. It was reported that FeOCl 
isotypic 
with TiNCl forms a similar intercalation compound with Py, and the electrical conductivity increases by about seven orders due to the charge transfer from the organic Lewis base to the FeOCl layers \cite{Kanamaru73}. The compound is a semiconductor, and does not become a superconductor. 

$n$-Alkyl monoamines (C$_n$H$_{2n+1}$NH$_2$, $3\leqq n\leqq12$) can form intercalation compounds with TiNCl, expanding the basal spacing to a value in the range of 12.0 to 37 \AA, with the alkyl chains oriented in various inclination angles to the layers. However, all of the compounds with $n$-alkyl monoamines were found to be not superconductors down to 2 K \cite{Yamanaka12}. It is interesting that ethylene diamine (NH$_2$CH$_2$CH$_2$NH$_2$) can form a similar intercalation compound with TiNCl with a basal spacing of 11.12 \AA, which 
surprisingly 
shows superconductivity with $T_c = 10.5$ K \cite{Yamanaka12}. Systematic studies have been done using alkylene diamines with different numbers of carbon atoms, NH$_2$C$_n$H$_{2n}$NH$_2$ ($2\leqq n\leqq12$) \cite{Yamanaka12}. Diamines with even number of carbon atoms appear to have larger superconducting 
volume fractions, and diamines with longer alkylene chains are favorable for higher $T_c$. The compound with $n=10$ (decamethylene diamine) shows a large superconducting 
volume fraction $>50$ \%, and $T_c = 17.1$ K, which is comparable to $T_c$'s of the alkali metal intercalated TiNCl. All kinds of the diamine molecules are intercalated between TiNCl layers with the alkylene chain axis parallel to the layers as shown in Figure 18. 

Mechanisms for the superconductivity with neutral amines are not clear, and remain open problems for physicists as well as chemists. It should be mentioned that the $\alpha$-structured HfNBr also forms similar intercalation compounds with alkali metals. The color of the crystals changed to black from pale yellow upon alkali metal intercalation. However, those were found to be insulators \cite{Yamanaka04}.

\section{Summary}

We have reviewed the layered nitride superconductors, especially 
the physical properties of the normal and superconducting 
states in the $\beta$-ZrNCl and HfNCl. On the basis of the above experimental and theoretical results, we summarized the physical properties of the layered nitride superconductors and compared them with other systems, as follows. 

\begin{enumerate}[(1)]
\item
The parent layered compounds are band insulators, which are changed into superconductors by electron-doping via intercalation of alkali, alkaline-earth, and rare-earth metals. The doping concentration can be continuously controlled in a wide range, and the basal spacing (interlayer distance) can be independently expanded by cointercalation of solvent molecules, keeping the doping concentration unchanged. 
\item
Reflecting the layered crystal structure, the electronic states are of highly two-dimensional character, 
whereas many superconductors based on band insulators have three-dimensional crystal and electronic structure. 
Experiments confirmed that 
the electronic properties are well described by the band 
calculations. 
Superconductivity emerges when carriers are doped into the 2D bands at the K and K$^\prime$ points in the hexagonal Brillouin zone. The carrier density as well as the density of states is 
low, 
as compared with other superconductors with similar values of $T_c$. 
\item
Superconducting 
phase 
exists next to an insulating state 
without any magnetic or electronic ordering, 
which is in marked contrast to superconductors with strong electron correlations, such as cuprates, pnictides, layered organics, and fullerides. 
\item
Anomalous doping dependence of $T_c$ is observed in Li$_x$ZrNCl, 
which makes this compound rare among other superconductors.  
$T_c$ 
does not show a dome-like dependence as in SrTiO$_3$, cuprates, and pnictides, and 
increases with decreasing carrier density toward superconductor-insulator boundary. 
Recently, a similar phase diagram has also been found in a completely different material, Sr$_{1-x}$La$_x$FBiS$_2$. 
\item
Anisotropy of the superconducting 
gap varies from isotropic to highly anisotropic with 
increasing 
doping, which is revealed by the specific heat and $\mu$SR experiments. 
\item
The enhancement of $T_c$ upon cointercalation of molecules, which modifies nesting properties of the Fermi surface,  gives indirect but crucial insights for the charge and/or spin fluctuation mechanisms of superconductivity. 
\item
Anisotropic superconducting 
gap may suggest unconventional pairing symmetry. $d+id$ pairing has been proposed as a possible pairing state by Hubbard-type models on the honeycomb lattice as in graphene. 
\item
Superconducting 
mechanism still 
remains mysterious.  
Unconventional mechanisms rather than typical phonon-mediated superconductivity can explain 
doping 
variation of superconducting 
gap 
and interlayer-distance dependence of $T_c$, but electron-phonon mechanism may be still relevant for explaining 
the high $T_c$. 
\item
The $\alpha$-structured compounds 
(TiNCl and TiNBr)  
also show superconductivity by electron doping via intercalation of alkali metals as well as organic bases. Upon cointercalation of solvent molecules with alkali metals, the basal spacing ($d$) of the compound increases adjusting to the molecular size. 
The $T_c$ increases linearly as a function of $1/d$, implying the importance of the Coulomb interlayer coupling in the pairing mechanism, 
in contrast to the $\beta$-structured compounds in which $T_c$ increases with the increase of $d$. 
\item
The coherence length along the $c$ direction $\xi_c$ of the $\alpha$-structured compounds is much larger than the interlayer spacing, while in the $\beta$-structured compounds $\xi_c$ is comparable to the basal spacing. These striking differences suggest that the superconducting 
layers of the $\beta$-structured compounds are presumably weakly Josephson coupled, while the $\alpha$-structured superconductors have more or less 3D characters. 
\end{enumerate}

These findings demonstrate the uniqueness of the layered nitride among superconductors. Although superconducting 
mechanisms are still under debate, their clarification is highly important and may provide a bridge between superconductivity in doped band insulator and strongly correlated systems.

\section*{Acknowledgements}
We would like to thank Y. Iwasa for useful comments and discussions. 
This work was partly supported by Grants-in-Aid for Specially Promoted Research (No. 2500003) and for Scientific Research from the Ministry of Education, Culture, Sports, Science, and the Technology of Japan (MEXT), the Funding Program for World-Leading Innovative R\&D on Science and Technology (FIRST Program) of the Japan Society for the Promotion of Science (JSPS), and SICORP-LEMSUPER EU-Japan Project No. 283214. 

\label{}





\end{document}